\documentclass[10pt,a4paper,onecolumn]{article}

\usepackage{achemso}
\usepackage[T1]{fontenc}       
\usepackage{graphicx}
\usepackage{dblfloatfix}
\usepackage{placeins}
\usepackage{authblk}
\usepackage{a4wide}
\usepackage{setspace}
\usepackage[margin=1in]{geometry}
\usepackage{physics}
\usepackage{color}
\usepackage{bm}
\usepackage{booktabs, tabularx} 
\newcolumntype{C}{>{\centering\arraybackslash}X}

\author[1]{Sanjoy Patra}
\author[1]{Amitav Sahu}
\author[1]{Vivek Tiwari \thanks{vivektiwari@iisc.ac.in}}
\affil[1]{Solid State and Structural Chemistry Unit, Indian Institute of Science, Bangalore, Karnataka 560012, India}

\begin{document}

\title {Trap Mediated Energy Transport via Vibronic Resonance}
\maketitle


\begin{abstract}
Controlling energy transfer through vibronic resonance is an interesting possibility. Exact treatment of non-adiabatic vibronic coupling is necessary to fully capture its role in driving energy transfer. However, exact treatment of vibrations in extended systems is expensive, sometimes requiring oversimplifying approximations to reduce vibrational dimensionality, and do not provide physical insights into which specific vibrational motions promote energy transport. Here we derive effective normal modes for excitonically coupled aggregates which reduce the overall high-dimensional vibronic Hamiltonian into independent one-dimensional Hamiltonians. Applying this approach on a trimer toy model, we demonstrate trap-mediated energy transport between electronically uncoupled sites. Bringing uncoupled sites into vibronic resonance converts the `trap’ into a `conduit’ for population transfer, while simultaneously minimizing trapped excitations. Visualizing energy transfer along the aggregate normal modes provides non-intuitive insights into which specific vibrational motions allow for trap-mediated energy transport by promoting vibronic mixing. 
\end{abstract}

\section{Introduction}
Role of vibrational motions in driving photoisomerization reactions\cite{Miller2015}, singlet exciton fission\cite{Kukura2015}, and energy and charge transfer\cite{Tiwari2013} in photosynthesis, is a subject of active investigation. Inducing specific vibrational motions can drive photoinduced charge transfer and phase transitions in molecular crystals\cite{Matsuzaki2006,Dawlaty2017,Frontiera2020}, and interfacial charge separation and electronic phase transitions in strongly correlated solids \cite{Prezhdo2016,Cavalleri2016}. The above examples motivate a need for identifying physically relevant vibrational motions which maximally promote electronic dynamics against mere spectator modes, in order to potentially control energy and charge transport by modulating vibronic interactions.\\
 
Excitonic dimer models\cite{Tiwari2013} have suggested that vibronic resonance between exciton energy gap and a quantum of vibrational excitation on the acceptor exciton causes strong non-adiabatic coupling of the resonant vibrational mode with the electronic subsystem, with potential relevance for enhancement\cite{Chin2014, Mancal2012, Thorwart2015, Lovett2014, Cao2015, Fuller2014,Grondelle2014,Dean2017} of energy and charge transfer. Vibronic resonances are likely\cite{JonasARPC} in condensed phase systems, and necessitate\cite{Tiwari2018,Tiwari2020} exact quantum treatment of resonant vibrations. However, vibronic resonances have a width, and incorporating several resonant vibrations in extended aggregates becomes computationally challenging. Moreover, incorporating several modes in the system does not distinguish promoter from spectator vibrational motions. \\


Here we formalize an effective-mode approach for excitonically coupled aggregates that allows us to exactly factor the overall high-dimensional energy transfer Hamiltonian into independent one-dimensional Hamiltonians. We identify aggregate spectator modes which do not participate in the vibronic mixing against normal modes which maximally promote energy delocalization. Applying this approach on a trimeric aggregate, we demonstrate vibronically enhanced population transfer between electronically \textit{uncoupled} donor-acceptor sites, which is mediated by a `trap'. Our approach highlights an interesting design principle of enhancing energy transport by modulating trap behavior through vibronic resonance.\\

\begin{figure*}[h!]
	\centering
	\includegraphics[width=3 in]{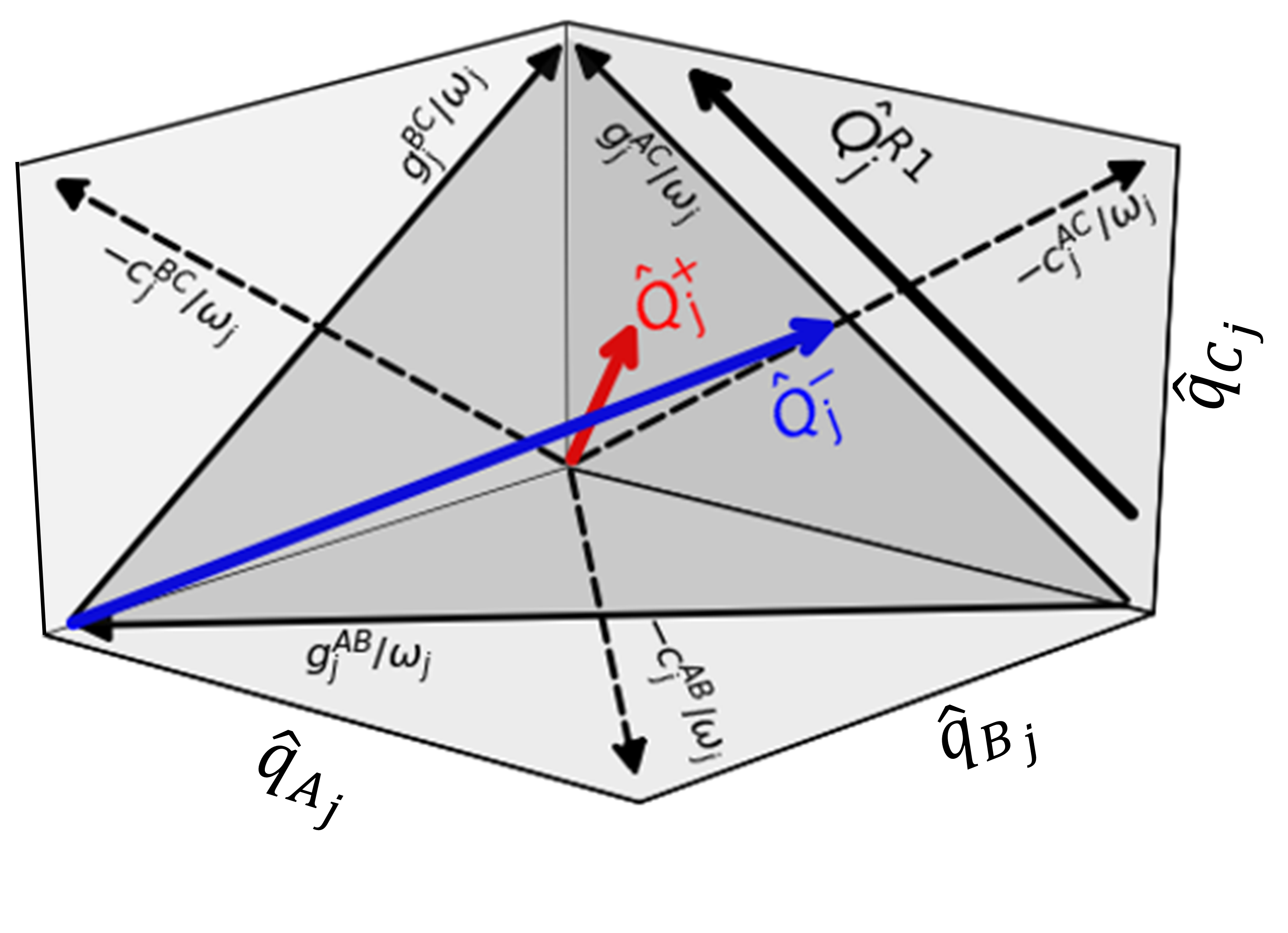}
	\caption{\footnotesize Effective normal modes for a trimeric aggregate. The $j^{th}$ set of intramolecular vibrational modes on each molecule, $\hat{q}_{A_j}$, $\hat{q}_{B_j}$ and $\hat{q}_{C_j}$, are represented along the three axes in a left-handed convention. The pairwise correlation ($\textbf{c}_j/\omega_j$) and energy gap tuning ($\textbf{g}_j/\omega_j$) coordinates are shown in thin dashed and solid arrows respectively. Normal modes of the aggregate are shown as thick solid vectors. $\hat{Q}_j^+$ normal mode of the trimer, shown in red, does not tune any energy gaps and is perpendicular to the plane formed by dimeric tuning vectors. $\hat{Q}_j^-$ normal mode of the trimer, shown in blue, tunes all nearest neighbor energy gaps, $A-B$ and $B-C$, lies in the plane formed by $A-B$ and $B-C$ tuning coordinates, and is orthogonal to the $A-C$ tuning coordinate by definition. Geometrically, it is seen that the remaining aggregate normal mode, $\hat{Q}_j^{R_1}$ is the $A-C$ tuning coordinate, and orthogonal to the normal modes $\hat{Q}_j^+$ and $\hat{Q}_j^-$.    
	}
	\label{fig:fig1}
\end{figure*}

For an excitonic dimer with molecules $A$ and $B$, motions\cite{Tiwari2017} along an energy gap tuning vector, $\textbf{g}^{AB}$ cause non-adiabatic vibronic mixing between excitons, and drive energy transfer. In contrast, motions along a correlation vector, $\textbf{c}^{AB}$ do not tune the inter-pigment excited state energy gap, and play no role in vibronic mixing. The aggregate normal modes are defined in terms of the pairwise tuning and correlation vectors between any pair of molecules $I$ and $J$ of the aggregate, $\textbf{c}^{IJ}$ and $\textbf{g}^{IJ}$ respectively, such that vibrational motions along the $j^{th}$ set of intramolecular vibrational mode on each molecule, $\{\hat{q}_{Ij}\}_{I=1}^N$, collectively contribute towards the global dynamics of the aggregate. The derivation is described in the Section S1 of the Supporting Information(SI), and briefly summarized in Figure \ref{fig:fig1}, using geometric considerations for a trimeric aggregate.  

\FloatBarrier

Figure \ref{fig:fig1}, plotted for the $j^{th}$ set of intramolecular vibrational modes of the trimer, shows the geometric relationship between the aggregate normal modes and the pairwise tuning and correlation modes of ref.\cite{Tiwari2017}. Along this mode, equal Franck-Condon (FC) displacement on all molecules, that is, $d_{A_j} = d_{B_j} = d_{C_j}$, is chosen for ease of visualization, although a general derivation is given in the Section S1, where
for each set $j$ of intramolecular vibrational modes, a $N$-dimensional Hamiltonian has been transformed to a sum of $N$ one-dimensional Hamiltonians. The key point is that the effect of each aggregate normal mode on vibronic mixing can now be treated \textit{independently} using only a one-dimensional Hamiltonian. As opposed to oversimplifications\cite{Tiwari2018,Tiwari2020} in $n$-particle approximations to the vibrational subspace, the aggregate normal modes treat vibronic coupling exactly, substantially reduce the vibrational space dimensionality, and deconvolute the overall vibronic mixing in terms of collective vibrational motions along individual aggregate normal modes. The starting point of our approach is similar to earlier effective-mode schemes\cite{Burghardt2005, Burghardt2019}. However, by not combining intramolecular modes of two different vibrational frequencies, allows reducing the overall Hamiltonian into \textit{independent} one-dimensional Hamiltonians along the aggregate normal modes, lending a physical interpretation of promoter versus spectator modes. \\

%

The trimer molecules are labeled as $A$, $B$ and $C$, in increasing order of site energy. The full trimer Hamiltonian is shown in Eqn.S1. The electronic part of the Hamiltonian can be written as --
\begin{equation}
\hat{H}_{el} =\left(%
\begin{array}{ccc}
-\Delta/2 & J & 0 \\
J & 0 & J \\
0 & J & \Delta/2 \\
\end{array}%
\right)
\label{eq9}
\end{equation}
The site energy gap between the molecules is $\Delta/2 = $177 cm$^{-1}$. Only the energetically nearest neighbors are directly coupled through Coulomb coupling $J =$ 66 cm$^{-1}$. The electronic coupling between the molecules lies in the weak to intermediate coupling regime\cite{Peterson1957}. We will treat explicitly two intramolecular vibrational modes per molecule of the trimer with frequencies 200 cm$^{-1}$ and 400 cm$^{-1}$, and small Huang-Rhys factors of 0.025, typical\cite{Bocian1995,Freiberg2011} for intramolecular vibrations in photosynthetic pigments. With $n_{vib}$ number of vibrational quanta for intramolecular vibration modes on each electronic state, the number of basis states scale rapidly as $3 n_{vib}^6$, making exact calculations of vibronic dynamics and nonlinear spectroscopic signatures computationally expensive.\\

The excitons resulting from Eqn.~\ref{eq9} are labeled as $\alpha$, $\beta$ and $\gamma$ in terms of increasing order of energy (Eqn. S22). Exciton $\alpha (\gamma)$ has less than 0.5\% character from molecule $C(A)$ such that molecules $A$ and $C$ have almost negligible direct electronic mixing between them. The Hamiltonian is set up such that $A-C$ electronic mixing is only mediated through an intermediate molecule $B$, with excitons $\alpha$($\gamma$) both having only $\sim$11\% $B$ character. \\

\begin{figure*}[h!]
	\centering
	\includegraphics[width=2.5 in]{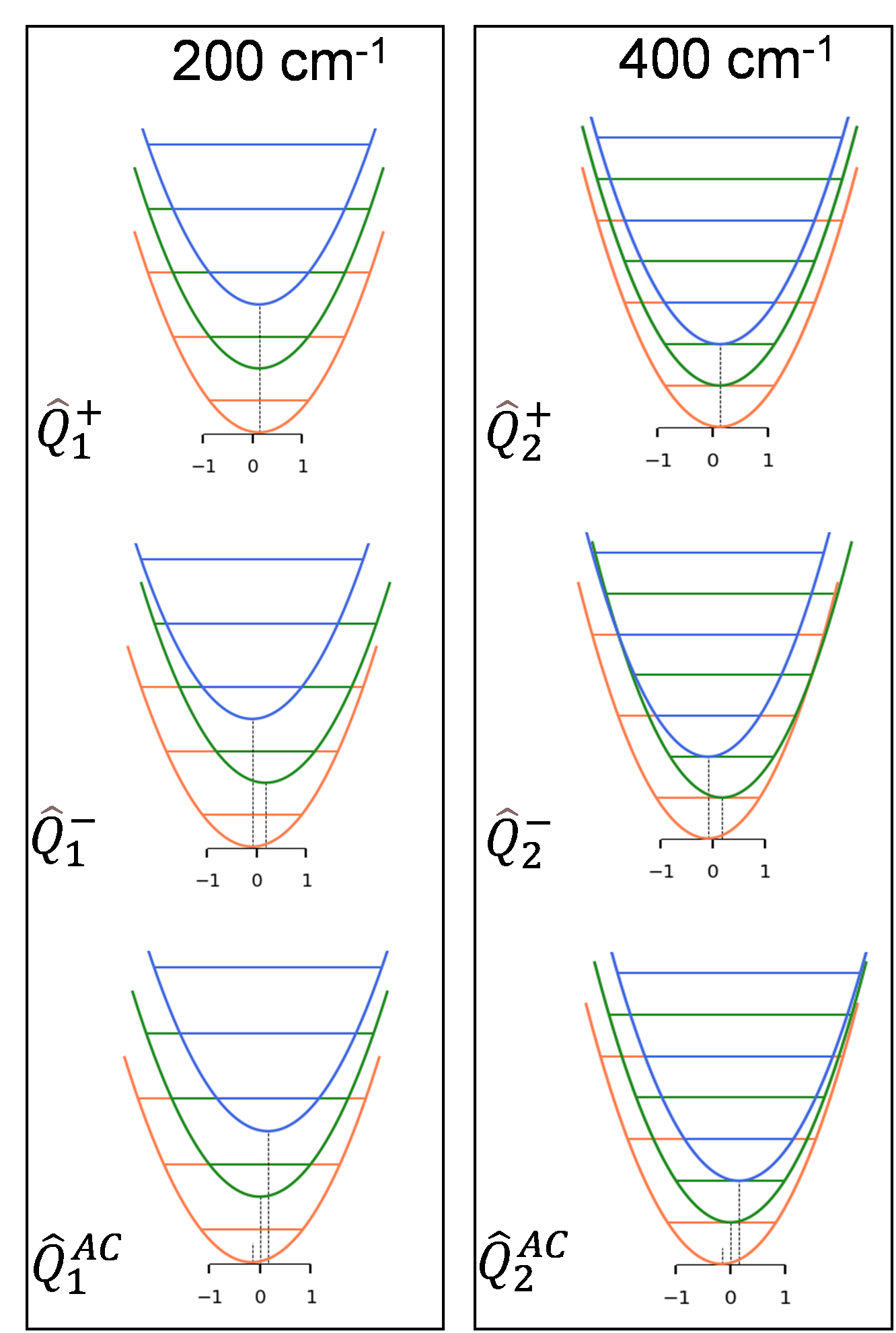}
	\caption{\footnotesize Singly-excited diabatic exciton states $\alpha$ (orange), $\beta$ (green) and $\gamma$ (blue) of the trimer plotted along respective one-dimensional aggregate normal mode slices in the 6-dimensional trimer Hamiltonian (Eqn. S1). The electronic Hamiltonian for the system is shown in Eqn.~\ref{eq9} and results in exciton energy gaps of 200 cm$^{-1}$ between successive excitons. The effective normal modes for the trimer, for each set $j$ of intramolecular vibrational modes, are labeled as $\hat{Q}_j^+$, $\hat{Q}_j^-$, and $\hat{Q}_j^{AC}$. The 200 cm$^{-1}$ mode is resonant between successive excitons resulting in resonant orange, green and blue vibrational levels. For the 400 cm$^{-1}$ mode, exciton $\beta$ is non-resonant, such that only orange and blue levels are resonant. The effective FC displacements in the diabatic excitonic basis are derived in Section S4 and listed in Tables S10 and S11. These are only a small fraction of the classical turning points of the zero-point levels, that is, $d \ll 1$ for the weakly coupled FC active vibrations considered here. All diabatic excitons have equal FC displacements along $\hat{Q}^+_j$ modes, such that motions along $\hat{Q}^+_j$ cause no change in relative energy gaps. In contrast, motions along $\hat{Q}^-_j$ tune relative energy gaps between excitons $\alpha-\beta$ and $\beta-\gamma$. Along $\hat{Q}^{AC}_j$, excitons $\alpha$ and $\gamma$ exhibit equal and opposite displacements, whereas exciton $\beta$ has zero FC displacement.}
	
	\label{fig:fig2}
\end{figure*}

The trimer normal modes (Figure \ref{fig:fig1}), are derived in Section S1 for each set $j=1,2$ of intramolecular modes. As geometrically seen in Figure \ref{fig:fig1}, $\hat{Q}_j^{R_1}$ is nothing but the dimeric $A-C$ tuning coordinate $\hat{Q}_{j}^{AC}$. As shown in Fig.~\ref{fig:fig2}, the two intramolecular vibrational frequencies, site energies and Coulomb couplings, have been chosen such that, along the 1$^{st}$ mode (200 cm$^{-1}$), a vibrational quantum on $\alpha$ is resonant with $\beta$, and a vibrational quantum on $\beta$ is resonant with $\gamma$. Thus, the intermediate site $B$ connects to both its nearest neighbors through vibronic resonance. Along the 2$^{nd}$ intramolecular mode (400 cm$^{-1}$), a vibrational quantum on exciton $\alpha$ is resonant with $\gamma$, while the intermediate site $B$ is not vibronically resonant with any other site. Overall, \textit{the trimer toy model is set up to study the questions -- \textbf{1.} can energy transfer between two uncoupled sites $A$ and $C$ be mediated through an intermediate site by leveraging vibronic resonance, \textbf{2.} which vibrational motions promote such vibronic mixing, \textbf{3.} whether it requires \textit{both} sites to be vibronically resonant with the intermediate molecule $B$ ?} \\

The above questions can be addressed by analyzing the role of 200 cm$^{-1}$ and 400 cm$^{-1}$ intramolecular vibrational modes in vibronic intensity borrowing seen in the absorption spectrum, and the corresponding population dynamics of the vibronic system Hamiltonian.\\

\begin{figure*}[h!]
	\centering
	\includegraphics[width=3.35 in]{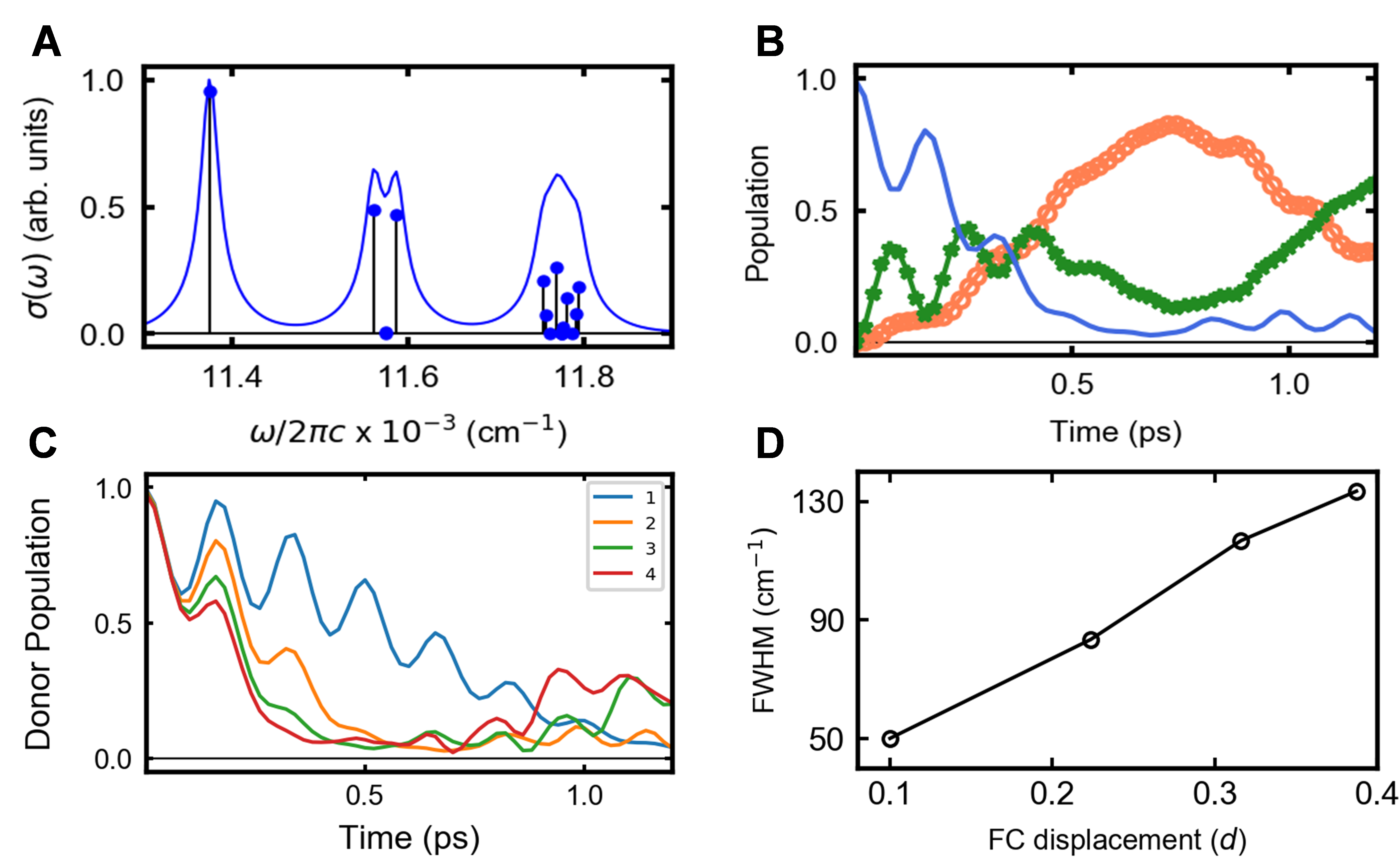}
	\caption{\footnotesize (\textbf{A})(Left) Exact calculation of absorption intensities of the trimer at 10 K, in the 6-dimensional intramolecular vibrational basis, with a 200 cm$^{-1}$ and 400 cm$^{-1}$ vibrational mode per molecule of the trimer. The absorption intensities are overlaid with normalized Lorentzian lineshapes and summarized in Table S3. (Right) Population on molecules $A$, $B$ and $C$, following excitation of molecule $C$ is shown as orange (\textbf{o}), green (\textbf{*}), and solid blue curves, respectively. \textbf{(B)} (Left) Population of donor molecule $C$ versus time, plotted for different excited state FC displacements along the intramolecular vibrational modes. (Right) Suppression of 200 cm$^{-1}$ purely electronic coherences as a function of the intramolecular FC displacement. The faster decay of electronic coherences is quantified as the increasing width of the 200 cm$^{-1}$ peak in the Fourier spectrum. More details of the analysis are provided in Figure S4.}
	\label{fig:fig3}
\end{figure*}

The trimer absorption spectrum and population dynamics resulting from calculations in the intramolecular vibrational basis are shown in Figure \ref{fig:fig3}A.
The transition strengths in the absorption spectra are calculated at 10 K. The line strengths are overlaid with normalized Lorentzian lineshapes to show that vibronic splittings of the order of a few wavenumbers can be easily overwhelmed by lineshape broadening. Following earlier\cite{Peters2017} wavepacket propagation approach, described briefly in the Section S5, allows us to visualize the dynamics of impulsively excited coherent superpositions of vibronic excitons. The initial wavepacket motions are purely dictated by the vibronic system Hamiltonian\cite{Burghardt2005}, and weaker vibronic system-bath couplings influence the dynamics on longer timescales. The vibronic splittings seen under the second exciton in Figure \ref{fig:fig3}A(left) arise due to resonant intensity borrowing from lowest $\beta$ exciton to three states on $\alpha$, each with a quantum of excitation along one of the three 200 cm$^{-1}$ intramolecular modes ($\hat{q}_{A_1}$, $\hat{q}_{B_1}$ and $\hat{q}_{C_1}$). Similarly, a total of 12 states are isoenergetic with the lowest $\gamma$ exciton. The resonant states in Figure S1 show that 3-particle basis states, where the vibrational excitation is allowed to be on two different sites apart from the electronically excited site, play an essential role in resonant vibronic mixing, and cannot be neglected even if the intermediate exciton is non-resonant. Using an analytic approach\cite{Tiwari2018,Tiwari2020}, Section S4 explains the intensity borrowing effects in the absorption spectrum.\\

Since exciton $\gamma$, which has $\sim$89\% $C$ character, is vibronically resonant with 12 other states (Figure S1), excitation of pigment $C$ leads to a rapid decay of population within $\sim$600 fs (Figure \ref{fig:fig3}A, right). As will become apparent in the analysis along the aggregate normal modes, purely electronic coherences oscillate with a period corresponding to the 200 cm$^{-1}$ exciton energy gaps and lead to only $\sim$34\% population transfer between directly coupled molecules $B-C$, and negligible transfer to acceptor site $A$ because of no electronic mixing between the donor-acceptor sites. In contrast, it is seen that vibronic resonance enhances population transfer to the uncoupled acceptor site $A$ from $<$5\% to >80\% within 700 fs, while bypassing transfer to the directly coupled intermediate site. Curiously, this is also accompanied by a rapid suppression of purely electronic coherences, \textit{without} introducing any static disorder or bath-induced decoherence. Since the vibronic system is not coupled to rest of the vibrational bath, the initial population decay from molecule $C$ is not irreversible, and shows recurrences beyond 1 ps. Figure \ref{fig:fig3}B shows that the suppression of electronic coherences due to vibronic resonance is directly related to the strength of linear vibronic coupling through the FC displacement of the intramolecular vibration which participates in resonant vibronic mixing. The suppression is ultimately caused by destructive interferences between the multiple coherent superpositions between different pairs of vibronically mixed states under the 2$^{nd}$ and 3$^{rd}$ exciton. Similar suppression of electronic coherences caused by vibronic resonances has been recently reported\cite{Makri2020_2} by Makri and co-workers in their path-integral calculations on bacteriochlorophyll aggregates. \textit{The above calculations in the intramolecular vibrational basis, although significantly more expensive, do not address questions regarding which specific vibrational motions, and along which of the two vibrational modes, promote vibronic mixing leading to intensity borrowing effects in the absorption spectrum, and >80\% population transfer between uncoupled molecules.} \\

\begin{figure*}[h!]
	\centering
	\includegraphics[width=6 in]{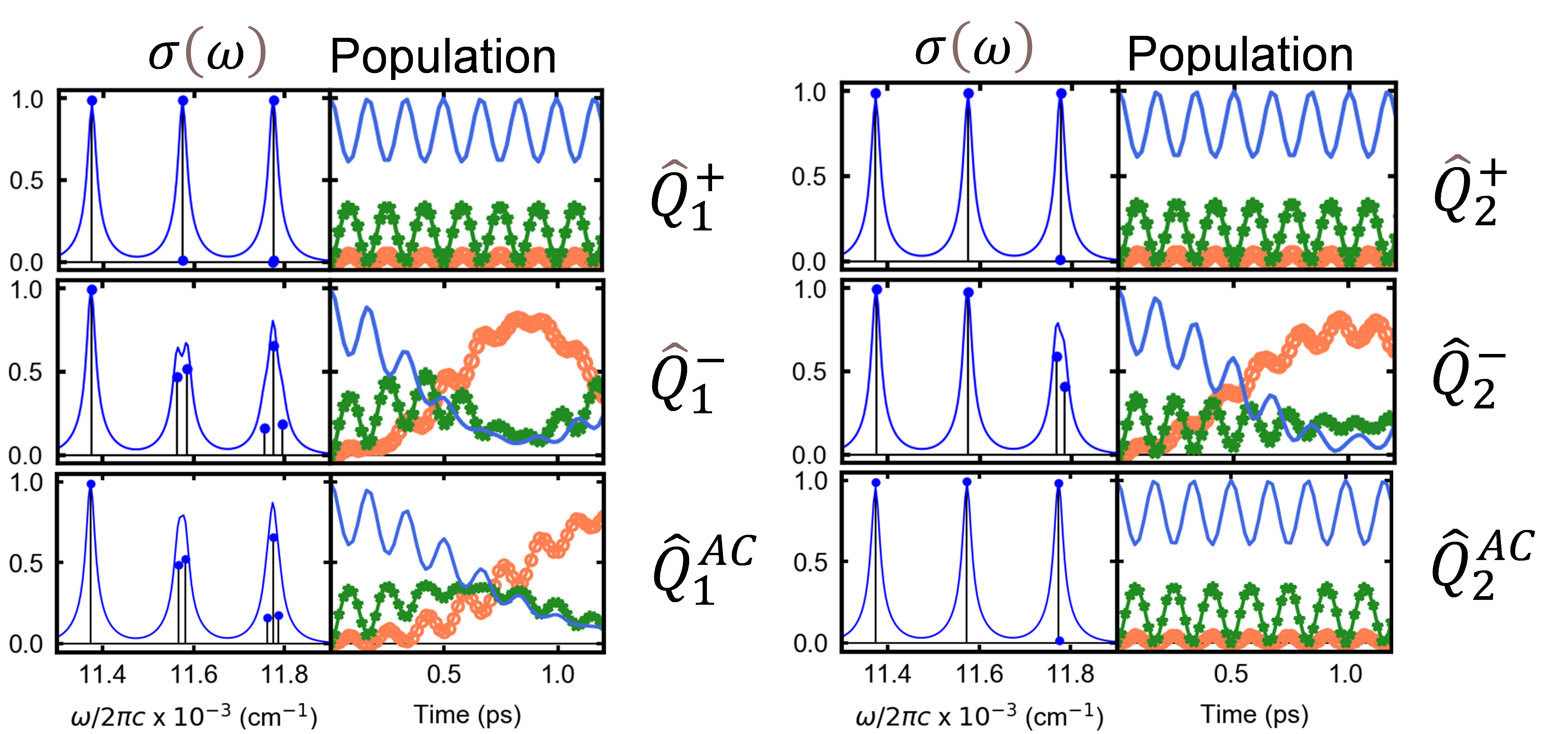}
	\caption{\footnotesize Exact calculation of the absorption intensities and population dynamics of the trimer at 10 K. Partitioning along aggregate normal modes $\hat{Q}_j^{+}$, $\hat{Q}_j^{-}$ and $\hat{Q}_j^{AC}$ reveals the specific vibrational motions which drive population transfer through vibronic mixing. The absorption intensities are overlaid with Lorentzian lineshapes. Population on molecules $A$, $B$ and $C$ is shown as orange (o), green (\textbf{*}), and solid blue curves, respectively. Aggregate modes with subscript `1' denote the 200 cm$^{-1}$ vibration. The set of modes with subscript `2' denote the 400 cm$^{-1}$ vibration. Along the aggregate correlation modes `+' which do not tune energy gaps, no vibronic splittings are seen in the absorption spectra and weak indirect electronic mixing between the molecules leads to <5\% $A-C$ population transfer. Vibronic mixing along the global tuning modes `-' and the remaining $A-C$ tuning modes, enhances $A-C$ population transfer to >80\%. Absorption peak strengths and locations are summarized in Tables S4-S9.}
	\label{fig:fig4}
\end{figure*}

Next, we show that the aggregate normal modes, which effectively represent one-dimensional slices in the six-dimensional potential energy surface, yield valuable physical insights into the above questions with significantly less expensive calculations. With $n_{vib}$ quanta along each aggregate mode, now the basis set size increases only as $3 \times n_{vib}$, significantly smaller than calculations in Figure \ref{fig:fig3}. Figure \ref{fig:fig4} (top row) shows the absorption spectra along the global correlation modes $\hat{Q}_1^{+}$ and $\hat{Q}_2^{+}$. The spectra show three peaks corresponding to each exciton, with a very weak 0-1 FC progression due to FC displacement $d \ll 1$. Motions along these modes do not mix the excitons, such that no intensity borrowing is seen. The corresponding wavepacket dynamics is purely adiabatic, and the amount of population transfer is dictated only by the weak electronic mixing in $\hat{H}_{el}$. Since molecules $A-C$ are not coupled directly, only <5\% $A-C$ population transfer occurs, and $\sim$34\% population transfer occurs between $B-C$. Dynamics along the `+' modes, is not dependent on vibronic resonance and is therefore identical along both 200 cm$^{-1}$ and 400 cm$^{-1}$ modes. The coherent oscillations arise due to purely electronic superpositions between excitons $\beta-\gamma$.\\

Aggregate modes which tune relative energy gaps show sharply contrasting spectra and dynamics. First considering the 200 cm$^{-1}$ vibration (panel A) which is resonant with energy gaps between successive excitons (Figure \ref{fig:fig2} left panel), motions along global tuning mode $\hat{Q}_1^{-}$, as well as $A-C$ tuning mode $\hat{Q}_1^{AC}$, both mix the excitons, resulting in vibronic splittings under excitons $\beta$ and $\gamma$ in the absorption spectra. Figure \ref{fig:fig2} shows that a 200 cm$^{-1}$ quantum of vibrational excitation on exciton $\alpha$, is resonant with $\beta$. The vibrational excitation on $\alpha$ mixes with $\beta$, and leads to two peaks of nearly equal intensities under $\beta$. In the same fashion, exciton $\gamma$ is resonant with two vibrational levels (seen as isoenergetic orange, green and blue levels in Figure \ref{fig:fig2}, left panel), resulting in total three vibronically mixed peaks under $\gamma$. Section S4 analytically explains the above absorption spectra. Larger vibronic coupling along $\hat{Q}_1^{-}$ results in larger vibronic splittings compared to those along $\hat{Q}_1^{AC}$. The vibronic intensity borrowing seen in the absorption spectra manifests in the corresponding population dynamics as vibronically enhanced population transfer, from only $\sim$5\% to >80\%, between molecules $A$ and $C$ which otherwise have negligible electronic mixing. Note that motions along both, $\hat{Q}_1^{-}$ and $\hat{Q}_1^{AC}$, lead to vibronic enhancement of $A-C$ population transfer.
Figures S2,S3 analyze a general case where $\alpha-\beta$ and $\beta-\gamma$ exciton energy gaps are not the same, such that two different vibrational modes, with simultaneous vibrational excitation along each of the two vibrational modes (Figure S2), is required to achieve direct $A-C$ population transfer. \textit{A pertinent question to ask is whether vibronic resonance between both, $\alpha-\beta$ and $\beta-\gamma$ is necessary for achieving population transfer between uncoupled sites $A$ and $C$ ?}\\

The above question is better understood when considering the 400 cm$^{-1}$ vibrational mode. Exciton $\beta$ with $\sim$78\% $B$ character, is not resonant with other excitons and hence no role in vibronically enhanced population transfer is expected. Comparing the effects of motions along $\hat{Q}_2^{-}$ and $\hat{Q}_2^{AC}$  on vibronic mixing and population transfer (middle and bottom figures in panel B of Figure \ref{fig:fig4}), it is seen that despite $\alpha-\gamma$ vibronic resonance, tuning $A-C$ energy gaps plays no role in vibronic mixing. The resulting absorption spectrum and population dynamics along $\hat{Q}_2^{AC}$ is nearly identical to that seen for the case of $\hat{Q}^{+}_{1,2}$ normal modes which do not tune any energy gaps. In contrast, enhancement of $A-C$ population transfer through $\alpha-\gamma$ vibronic resonance is caused by motions along $\hat{Q}_2^{-}$, which \textit{do not} tune the $A-C$ energy gap (see Figure \ref{fig:fig2}). 

\begin{figure*}[h!]
	\centering
	\includegraphics[width=3.35 in]{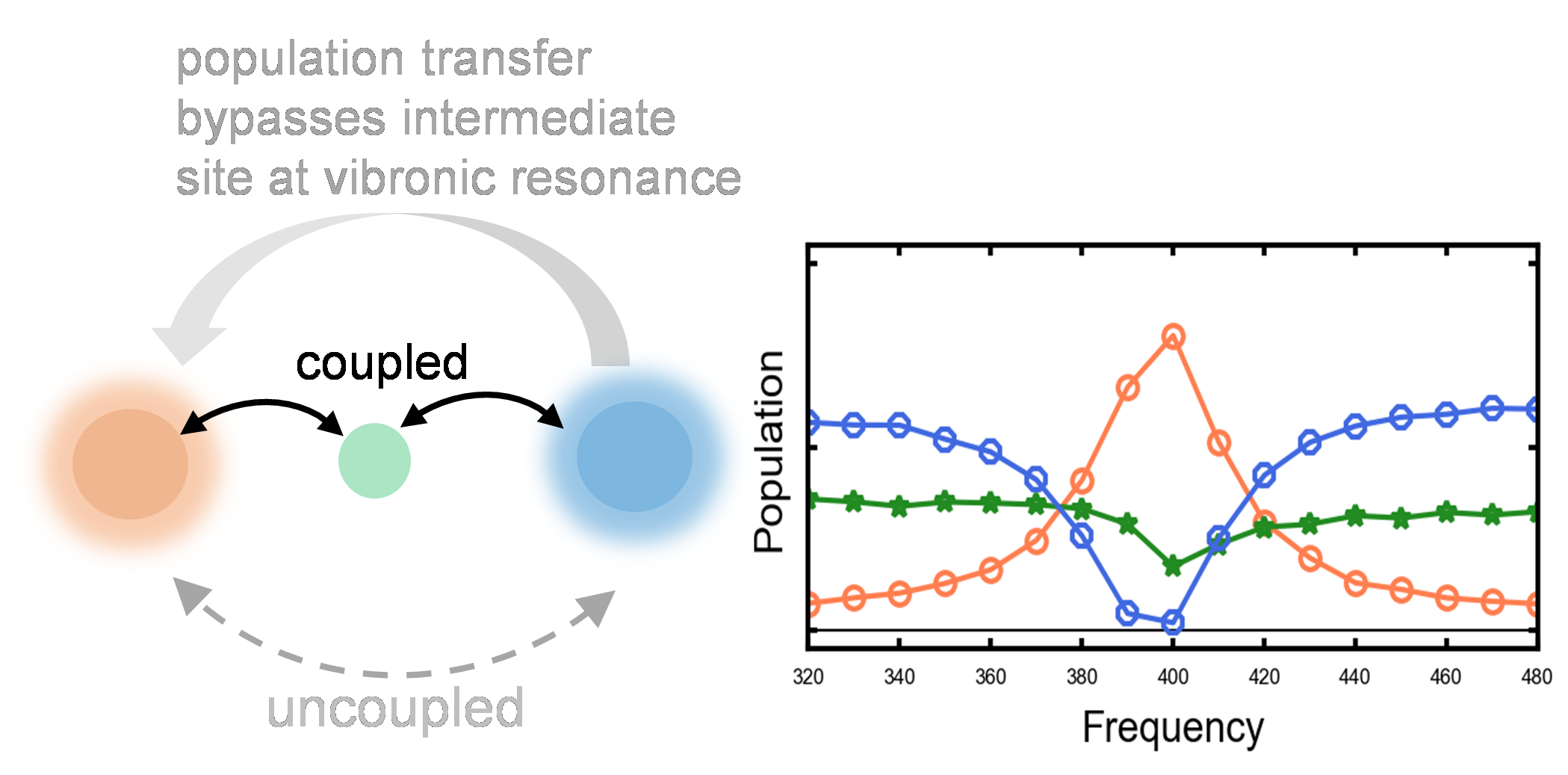}
	\caption{\footnotesize Trap-mediated energy transfer at vibronic resonance driven by global tuning coordinate $\hat{Q}^{-}_2$. (\textbf{Left}) The donor (blue) and acceptor (orange) sites are electronically uncoupled, while the `trap' site (green) mixes with both such that donor(acceptor) excitons $\sim$11\% character from the trap site, while only $\sim$0.5\% character from the acceptor(donor) site. (\textbf{Right}) Minimum population on donor site plotted as a function of intramolecular vibrational frequency. Population on the acceptor and trap sites are plotted for the time point at which the donor population is minimum. Away from vibronic resonance, only incomplete population transfer occurs between donor/trap sites, and negligible population is transferred from the trap to the acceptor site. As the donor/acceptor excitons are brought into vibronic resonance, while the trap exciton remains non-resonant, donor population transfer is enhanced to $\sim$100\%, such that the population is directly transferred to the acceptor while transfer to the trap is minimized.}
	\label{fig:fig5}
\end{figure*}
\FloatBarrier  

The physical implications of the above, somewhat counter-intuitive result can be appreciated if the intermediate molecule $B$ is envisaged as a `trap' state which is electronically coupled to mutually uncoupled donor-acceptor sites, $C$ and $A$, respectively. Figure \ref{fig:fig5} shows the population on the donor, acceptor and trap sites as a function of intramolecular vibrational frequency, which is scanned across the $\alpha-\gamma$ exciton energy gap. Away from vibronic resonance, the trap receives $\sim$34\% excitation from the excited donor due to direct electronic coupling between them, while very less population is transferred to the acceptor because of negligible electronic mixing between the donor and acceptor sites. However, as the donor-acceptor excitons ($\gamma$ and $\alpha$, respectively) are brought into vibronic resonance, the `trap' becomes a `conduit' to enhance population transferred from the donor, from $\sim$34\% to $\sim$100\%. Figure \ref{fig:fig5}(right) shows that all of this extra donor population is transferred to the acceptor site, while transfer to the trap is simultaneously minimized. Importantly, in order to act like a `conduit', the trap can remain non-resonant with the donor and acceptor excitons. The aggregate normal modes derived earlier allow for the analysis in Figure \ref{fig:fig4}, and identify the specific vibrational motions which allow the `trap' to become a `conduit'. It is not the $A-C$ tuning motions, but rather the relative donor-trap and trap-acceptor energy gap tuning motions along the global tuning mode $\hat{Q}^{-}$ which drive this dynamics. The role of $\hat{Q}^{-}$ versus $\hat{Q}^{AC}$ is explained analytically in Section S4 by considering the electronically off-diagonal vibronic coupling matrix elements. \\  

Analyzing vibronic mixing contributions along the aggregate normal modes for a trimer highlights interesting design principles for inducing energy transfer between two uncoupled sites through an intermediate site. First, for vibrational modes which are successively resonant between the acceptor, intermediate and donor excitons, both global energy gap tuning motions, as well as second-nearest neighbor tuning motions promote energy transfer. Second, even a non-resonant intermediate trap which mixes with the two uncoupled sites can mediate energy transfer. Vibronic resonance converts the trap into a conduit to promote direct energy transfer between the two sites, which otherwise do not exchange energy due to negligible electronic mixing. \\

%
By analyzing the problem along the aggregate normal modes, we have identified the collective vibrational motions which maximally promote vibronic mixing, against spectator modes which need not be explicitly treated in the system Hamiltonian. The reduction in dimensionality in our approach is exact in nature, and does not require oversimplifying\cite{Tiwari2018,Tiwari2020} $n$-particle approximations. This presents an attractive route towards simulations\cite{CinaARPC,Mancal2016, Mukamel1998} of wavemixing pathways in two-dimensional electronic spectroscopy on larger aggregate systems, with exact treatment of vibronic mixing. Future extensions of our approach using a bath decoherence function\cite{Tiwari2017}, or including bath-specific relaxation\cite{Ishizaki2009} in hierarchical quantum master equations, can ascertain the effect of fluctuating environments on the collective vibrational modes which maximally promote energy transfer in extended aggregates.\\

\section{Acknowledgments}
SP and AS acknowledge research fellowship from the Indian Institute of Science (IISc). VT acknowledges IISc startup grant number SG/MHRD-18-0020. This project is supported by Department of Atomic Energy, India under grant sanction number 58/20/31/2019-BRNS, and by Science and Engineering Research Board, India under grant sanction number CRG/2019/003691.

\bibliography{VTlibrary}
\bibliographystyle{unsrt}

\end{document}